\documentclass[11pt]{cernrep}
\usepackage{graphicx}
\usepackage{cite,./mcite}
\begin{document}
\title{Diffractive Physics in ALICE \\ Proceedings Workshop HERA and
  the LHC, may 26-30, 2008, CERN}
\author{R. Schicker}
\institute{Physikalisches Inst., Philosophenweg 12, 69120 Heidelberg}
\maketitle
\begin{abstract}
The ALICE detector at the Large Hadron Collider (LHC) consists of a
central barrel, a muon spectrometer and neutron calorimeters at $0^o$.
Additional detectors for event classification and for trigger purposes 
are placed on both sides of the central barrel. Such a  geometry allows
the definition of a diffractive gap trigger by requiring no activity
in the additional detectors. I discuss some physics topics which become
accessible by this gap trigger.

\end{abstract}

\section{The ALICE Experiment}

The ALICE experiment is presently being commissioned at the
Large Hadron Collider (LHC)\cite{Alice1,Alice2}. The ALICE experiment 
consists of a central barrel covering the pseudorapidity 
range $-0.9 < \eta < 0.9$ and a muon spectrometer in the 
range $-4.0<\eta<-2.4$. Additional detectors for trigger purposes and 
for event classification exist in the 
range $ -4.0 < \eta < 5.0 $. The ALICE physics program
foresees data taking in pp  and PbPb collisions at 
luminosities up to $\cal{L}$ = $5\times 10^{30}cm^{-2}s^{-1}$ and 
$\cal{L}$ = $10^{27}cm^{-2}s^{-1}$, respectively. 
An asymmetric system pPb will be measured at a luminosity of
$\cal{L}$ = $10^{29}cm^{-2}s^{-1}$.

\subsection{The ALICE Central Barrel}

The detectors in the ALICE central barrel track and identify 
hadrons, electrons and photons in the pseudorapidity range 
$ -0.9 < \eta < 0.9$. The magnetic field strength 
of 0.5 T  allows the measurement of tracks from very low transverse 
momenta  of about 100 MeV/c to fairly high values of about 100 GeV/c. 
The tracking detectors are designed to reconstruct secondary vertices 
resulting from decays of hyperons, D and B mesons. The granularity of
the central barrel detectors is chosen such that particle tracking and
identification can be achieved in a high multiplicity environment of
up to 8000 particles per unit of rapidity. The main detector systems
for these tasks are the Inner Tracking System, the Time Projection
Chamber, the Transition Radiation Detector and the Time of Flight
array. These systems cover the full azimuthal angle within the 
pseudorapidity range $ -0.9 < \eta < 0.9$ and are described below. 
Additional detectors with partial coverage 
of the central barrel are a PHOton Spectrometer (PHOS), an
electromagnetic calorimeter (EMCAL) and  a High-Momentum Particle 
Identification Detector (HMPID). 

\subsubsection{The Inner Tracking System}

The Inner Tracking System (ITS) consists of six cylindrical layers of
silicon detectors at radii from 4 cm to 44 cm. The minimum radius
is determined by the beam pipe dimensions whereas the maximum radius 
chosen is determined by the necessity of efficient track matching with
the outer detectors in the central barrel. 
The innermost layer extends over the range 
$ -2 < \eta < 2 $ such that there is continous overlap with 
event classification detectors outside of the central barrel.
Due to the high particle density of up to 80 particles/cm$^{2}$ and 
in order to achieve the required tracking resolution, pixel detectors 
have been chosen for the first two layers. Silicon drift detectors 
are located in the  middle two layers whereas double sided silicon
strip detectors are in the outer two layers. 
   
\subsubsection{The Time Projection Chamber}

The Time Projection Chamber (TPC) is the main tracking detector in the 
central barrel. The inner and outer radii of the active volume are 
84.5 cm and 246.6 cm, respectively. The full radial track length is 
measured in the pseudorapidity range $ -0.9 < \eta < 0.9$ whereas
tracks with at least one third of nominal radial length  are covered in the 
pseudorapidity range $ -1.5 < \eta < 1.5$. Particle identification
is achieved by measuring the specific ionization loss.    
The chosen geometry results in a drift time of about 90 $\mu$s. This
long drift time is the factor limiting the proton-proton luminosity
to the value mentioned above.   
   
\subsubsection{The Transition Radiation Detector}

The principal goal of the Transition Radiation Detector (TRD) is to
provide electron identification in the momentum range larger than 
1 GeV/c. In this range, the electron identification by energy
loss in the TPC is no longer sufficient. Since the TRD is a
fast tracker, the TRD information can be used for an efficient trigger 
on high transverse  momentum electrons. In addition, the position 
information from the TRD system improves the tracking performance of 
the central barrel. 

\subsubsection{The Time of Flight Detector}

The Time-Of-Flight (TOF) array is located at a radial distance 
from 3.7 m to 4.0 m. 
The TOF information is used for particle identification in the
range 0.2 GeV/c $ < p_{T} < $ 2.5 GeV/c. 
For this detector, the Multi-gap Resistive-Plate (MRPC) technology was 
chosen. A  strip with an active area of 120x7.4 cm$^{2}$ consists 
of pads of 3.5 cm length and 2.5 cm width.  
  
\subsubsection{The Central Barrel Performance}

The ITS, TPC and TRD detectors described above are the main tracking 
detectors in the central barrel. With the information from these 
detectors, particles with momenta as low as 100 MeV/c can be tracked.

\begin{figure}[h]
\begin{minipage}[t]{8.0cm}
\includegraphics*[scale=0.40]{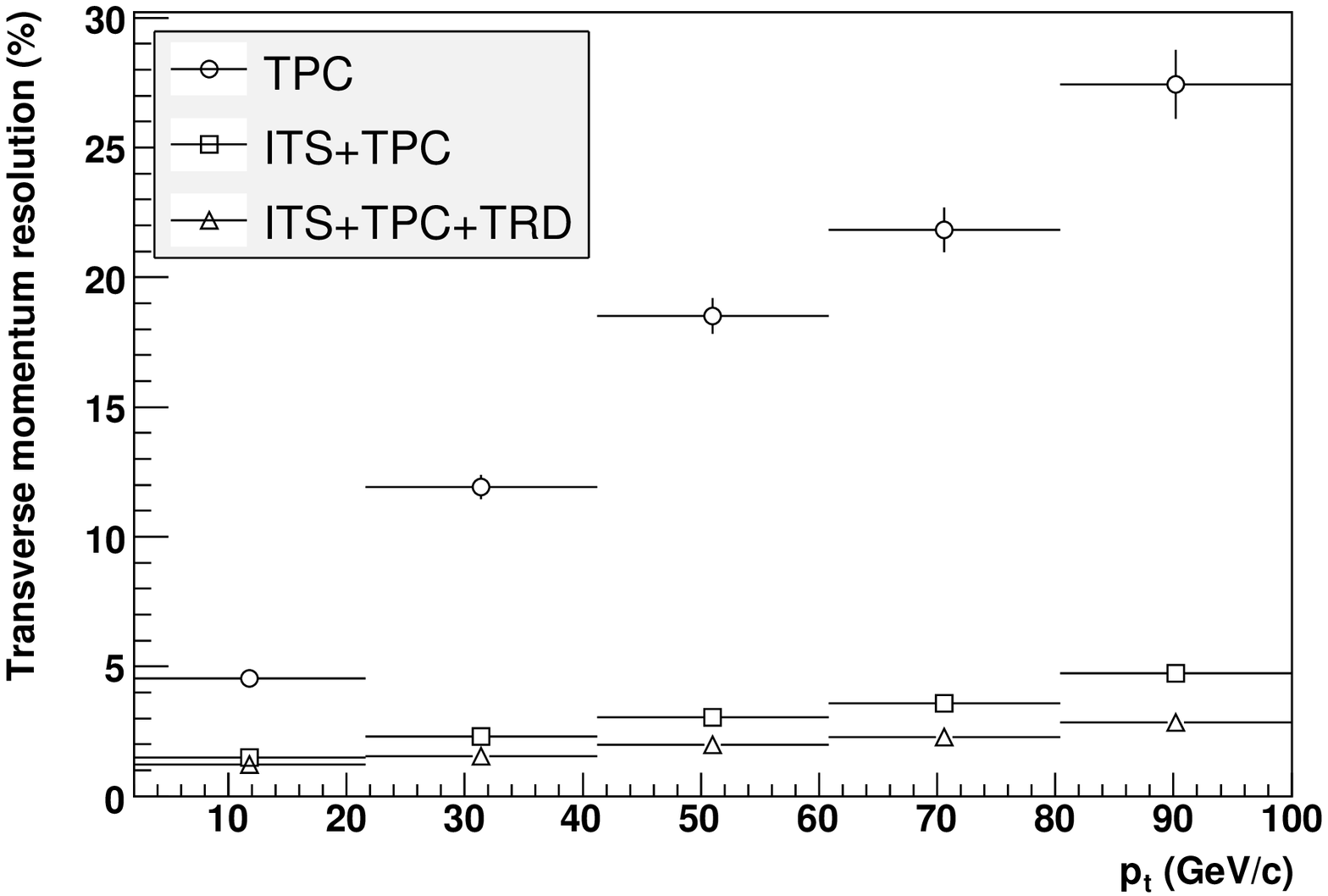}
\caption{Central barrel tracking resolution}
\label{fig:trl_perf}
\end{minipage}\hfill
\begin{minipage}[t]{8.0cm}
\includegraphics*[scale=0.43]{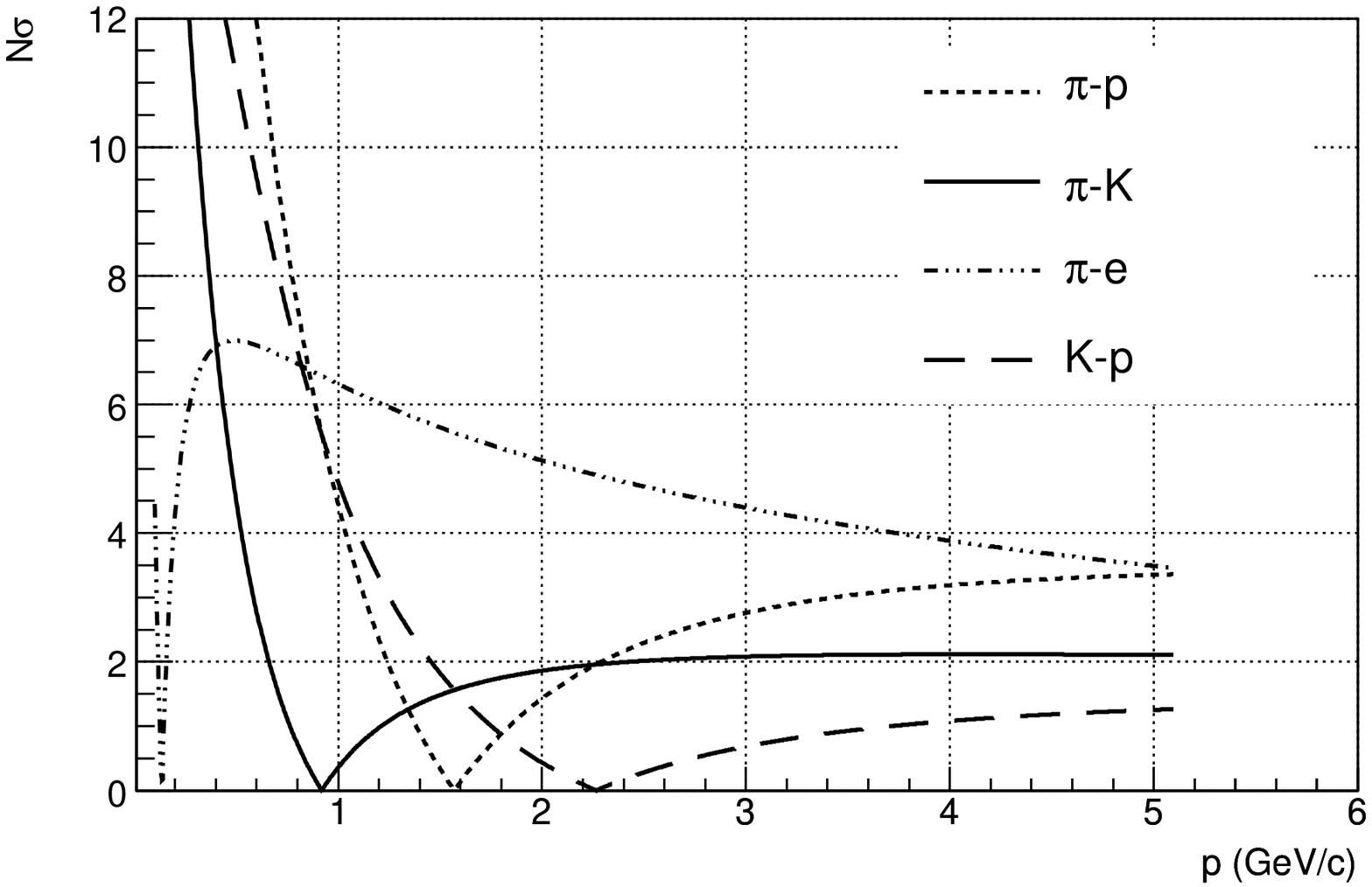}
\caption{Particle identification by dE/dx measurement}
\label{fig:dedx_pid}
\end{minipage}
\end{figure}

Fig.\ref{fig:trl_perf} shows the transverse momentum resolution
as expected from simulations. The TPC alone achieves a resolution of
approximately 3\% at a transverse momentum of $p_{T}$ = 10 GeV/c.
Adding the information from ITS and TRD on the inner and outer side,
respectively, improves the resolution considerably due to the increased
leverage. The combined transverse momentum  resolution from the ITS, TPC and
TRD detector is expected to be about 3\% at a transverse momentum
of $p_{T}$ = 100 GeV/c.

Particle identification is achieved in the central barrel by 
different methods. The specific energy loss is
measured by the TPC, the TRD and
the strip and drift detectors of the ITS.
Fig.\ref{fig:dedx_pid} shows the combined particle identification 
capability  by dE/dx measurement as a function of momentum.  The 
separation of different particle species is shown in units of the 
resolution of the dE/dx measurement. 
The electron-pion separation at high momenta is significantly improved 
by the information of the TRD system.

\subsection{The ALICE Zero Degree Neutron Calorimeter}

The Zero Degree Neutron Calorimeters (ZDC) are placed on both sides of the 
interaction point at a distance of 116 m\cite{ZDC}. The
ZDC information can be used to select different diffractive
topologies. Events of the type $pp \rightarrow ppX$ do not deposit energy
in these calorimeters, events $pp \rightarrow pN^{*}X$ will have
energy in one of the calorimeters whereas events 
$pp \rightarrow N^{*}N^{*}X$ will have energy deposited in both calorimeters. 
Here, X denotes a centrally produced diffractive state from which the 
diffractive L0 trigger is derived as described below. 

\section{The ALICE diffractive gap trigger}

Additional detectors for event classification and trigger purposes
are located on both sides of the ALICE central barrel. First, an array
of scintillator detectors (V0) is placed on both sides of the 
central barrel. These arrays are labeled V0A and V0C on the 
two sides, respectively. Each of these arrays covers a pseudorapidity
interval of about two units with a fourfold segmentation of half a 
unit. The azimuthal coverage is divided into eight 
segments of 45$^{0}$ degrees hence each array is composed  of 32
individual counters.  
Second, a Forward Multiplicity Detector (FMD) is located on both sides 
of the central barrel. The pseudorapidity coverage of this detector
is $-3.4 < \eta < -1.7$ and $1.7 < \eta < 5.1$, respectively.

\begin{figure}[htb]
\begin{center}    
\includegraphics*[scale=0.34]{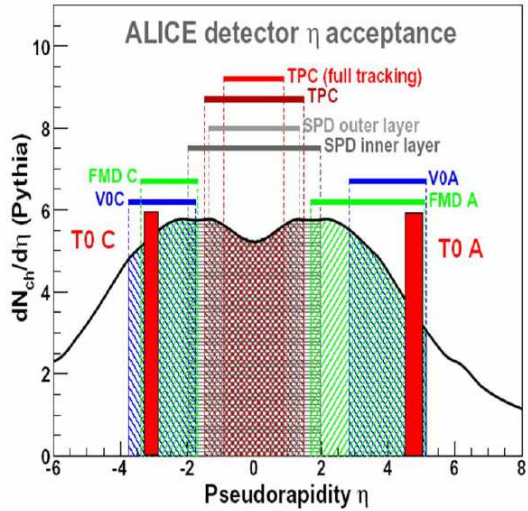}
\caption{Pseudorapidity coverage of trigger detectors and of detectors
in central barrel}
\label{fig:acc}
\end{center}
\end{figure}

Fig.\ref{fig:acc} shows the pseudorapidity coverage of the detector
systems described above. The geometry of the ALICE central barrel 
in conjunction with the additional detectors V0 and FMD is well suited
for the definition of a rapidity gap trigger. The ALICE trigger system
consists of a Central Trigger Processor (CTP) and is designed as a 
multi-level scheme with L0,L1 and L2 levels and a 
high-level trigger (HLT). 
A rapidity gap trigger can be defined by the requirement of signals 
coming from the central barrel detectors while V0 and FMD not showing 
any activity. Such a scheme requires a trigger signal from within 
the central barrel for L0 decision. The pixel detector of the ITS 
system is suited for delivering such a signal\cite{pixel}. 
Alternatively, this L0 signal can be derived from the TOF detector.

The high level trigger HLT has access to the information of all the 
detectors shown in Fig.\ref{fig:acc} and will hence be able to select
events with rapidity gaps in the range $-4 < \eta < -1$ and 
$1 < \eta < 5$. These gaps extend over seven units of pseudorapidity
and are hence expected to suppress minimum bias inelastic events
by many orders of magnitude.   

In addition to the scheme described above, the ALICE diffractive L0 
trigger signal can be generated from the Neutron ZDC if no central state 
is present in the reaction. A L0 signal from ZDC does, however,
not arrive at the CTP within the standard L0 time window. A L0 trigger
from ZDC is therefore only possible during special data taking runs 
for  which the standard L0 time limit is extended. 
The possibility of such data taking is currently under discussion.     

\section{ALICE diffractive physics}

The tracking capabilities at very low transverse momenta in
conjunction with the excellent particle identification make ALICE an 
unique facility at LHC to pursue a long term physics program of
diffractive physics. The low luminosity of ALICE
as compared to the other LHC experiments restricts the ALICE physics 
program to reactions with cross section at a level of a few nb per unit 
of rapidity.

\begin{figure}[htb]
\begin{center}
\includegraphics*[scale=0.28]{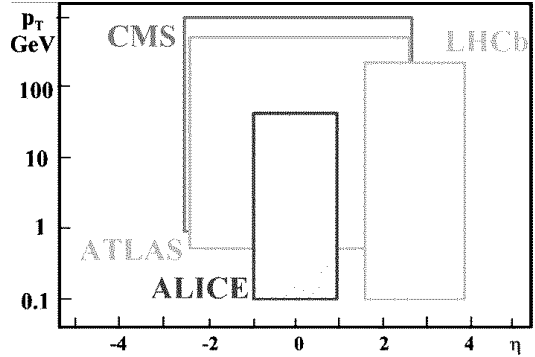}
\caption{Rapidity and transverse momentum acceptance of the LHC experiments}
\label{fig:acc_all}
\end{center}
\end{figure}

Fig.\ref{fig:acc_all} shows the transverse momentum acceptance of the 
four main LHC experiments. Not shown in this
figure is the acceptance of the TOTEM experiment which has a
physics program of measurements of total cross section, elastic
scattering and soft diffraction\cite{TOTEM}. The acceptance of the
TOTEM telescopes is in the range of $ 3.1 <  | \eta |  < 4.7$ and 
$5.3 < | \eta | <6.5$. 
The CMS transverse momentum acceptance of about 1 GeV/c shown
in Fig.\ref{fig:acc_all} represents a nominal value. The CMS analysis 
framework foresees the reconstruction of a few selected data samples to
values as low as 0.2 GeV/c\cite{CMS}.

\section{Signatures of Pomeron}
\label{sec:pomeron}

The geometry of the ALICE experiment is suited for measuring a centrally
produced diffractive state with a rapidity gap on either side. Such a 
topology can result, among other, from double Pomeron exchange with 
subsequent hadronization of the central state. It is expected that the
secondaries from  Pomeron-Pomeron fusion events show markedly 
different characteristics as compared to secondaries from inelastic
minimum bias events.

First, it is expected that the production cross section of glueball states
in Pomeron fusion is larger as compared to inelastic minimum bias
events. It will therefore be interesting to study the resonances produced
in the central region when two rapidity gaps are required\cite{close}. 

Second, the slope $\alpha'$ of the Pomeron trajectory is rather small:
$\alpha' \sim$ 0.25 GeV$^{-2}$ in DL fit and  $\alpha' \sim$ 0.1
GeV$^{-2}$ in vector meson production at HERA\cite{DL}. These values of
$\alpha'$ in conjunction with the small t-slope ($<$ 1 GeV$^{-2}$ ) 
of the triple Pomeron vertex indicate that the mean transverse
momentum $k_t$ in the Pomeron wave function is relatively large 
$\alpha' \sim$ 1/$k_t^2$, most probably \mbox{$k_t >$ 1 GeV}. The transverse 
momenta of secondaries produced in Pomeron-Pomeron interactions are of 
the order of this $k_t$. Thus the mean transverse momenta of secondaries 
produced in Pomeron-Pomeron fusion is expected to be larger as compared to 
inelastic minimum bias events. 

Third, the large $k_t$ described above corresponds to a large 
effective temperature. A suppression of strange quark production is 
not expected. Hence the K/$\pi$ ratio is expected to be enhanced in 
Pomeron-Pomeron fusion as compared to inelastic minimum bias 
events\cite{akesson}. Similarly, the $\eta$/$\pi$ and
$\eta'$/$\pi$ ratios are 
expected to be enhanced due to the hidden strangeness content and due 
to the gluon components in the Fock states of $\eta,\eta'$.

\section{Signatures of Odderon}
\label{sec:odderon}

The Odderon was first postulated in 1973 and is represented  
by color singlet exchange with negative C-parity\cite{nicolescu}. Due to its
negative C-parity, Odderon exchange can lead to differences between
particle-particle and particle-antiparticle scattering. In QCD, 
the Odderon can be a three gluon object in a symmetric color state.
Due to the third gluon involved in the exchange, a suppression by the 
coupling $\alpha_s$ is expected as compared to the two gluon Pomeron
exchange. However, finding experimental signatures of the Odderon 
exchange has so far turned out to be extremely difficult\cite{ewerz}.
A continued non-observation of Odderon signatures 
would put considerable doubt on the formulation of high energy
scattering by gluon exchange\cite{pomeron_qcd}. The best evidence 
so far for Odderon 
exchange was established as a difference between the differential
cross sections for elastic $pp$ and $p\bar{p}$ scattering 
at $\sqrt{s}$ = 53 GeV at the CERN ISR. The $pp$ cross section
displays a dip at t = -1.3 GeV$^2$ whereas the $p\bar{p}$ cross
section levels off. Such a behaviour is typical for negative
C-exchange and cannot be due to mesonic Reggeons only.   

\subsection{Signatures of Odderon Cross Sections}

Signatures of Odderon exchanges can be looked for in exclusive 
reactions where the Odderon (besides the Photon) is the only possible
exchange. Diffractively produced C-even states such as pseudoscalar
or tensor mesons can result from Photon-Photon, Photon-Odderon and
Odderon-Odderon exchange. Any excess measured beyond the well
understood Photon-Photon contribution would indicate an Odderon
contribution.

Diffractively produced C-odd states such as vector mesons 
$\phi, J/\psi, \Upsilon$ can result from Photon-Pomeron or 
Odderon-Pomeron exchange. Any excess beyond the Photon contribution
would be indication of Odderon exchange. 

Estimates of cross section for diffractively produced $J/\psi$ in pp
collisions at LHC energies were first given by Sch\"{a}fer et 
al\cite{schaefer}. More refined calculations by Bzdak et al result in 
a t-integrated photon contribution of $\frac{d\sigma}{dy}\mid_{y=0} \;
\sim$ 15 nb and a t-integrated Odderon contribution of   
$\frac{d\sigma}{dy}\mid_{y=0} \; \sim$ 1 nb\cite{bzdak}. 
These two numbers carry large uncertainties, the upper and lower 
limit of these numbers vary by about an order of magnitude. This cross 
section is, however, at a level where in 10$^6$ s of ALICE data taking the 
$J/\psi$ can be measured in its e$^+$e$^-$ decay channel at a level 
of 4\% statistical uncertainty. 

\begin{figure}[htb]
\begin{center}
\includegraphics*[scale=0.68]{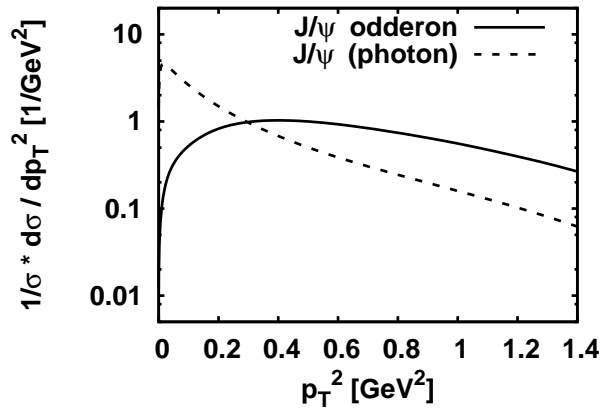}
\caption{The J/$\psi$ transverse momentum distribution for
the photon and Odderon contributions}
\label{fig:odderon_pt}
\end{center}
\end{figure}

Due to the different t-dependence, 
the Photon and Odderon contribution result in different transverse
momentum distribution $p_T$ of the $J/\psi$. 
The photon and Odderon contributions are shown in Fig.\ref{fig:odderon_pt} 
by the dotted and solid lines, respectively. A careful transverse 
momentum analysis of the $J/\psi$ might therefore allow to disentangle
the Odderon contribution. 

\subsection{Signatures of Odderon Interference Effects}

If the diffractively produced final state is not an eigenstate of
C-parity, then interference effects between photon-Pomeron and
photon-Odderon amplitudes can be analyzed. 

\begin{figure}[htb]
\begin{center}
\includegraphics*[scale=1.0]{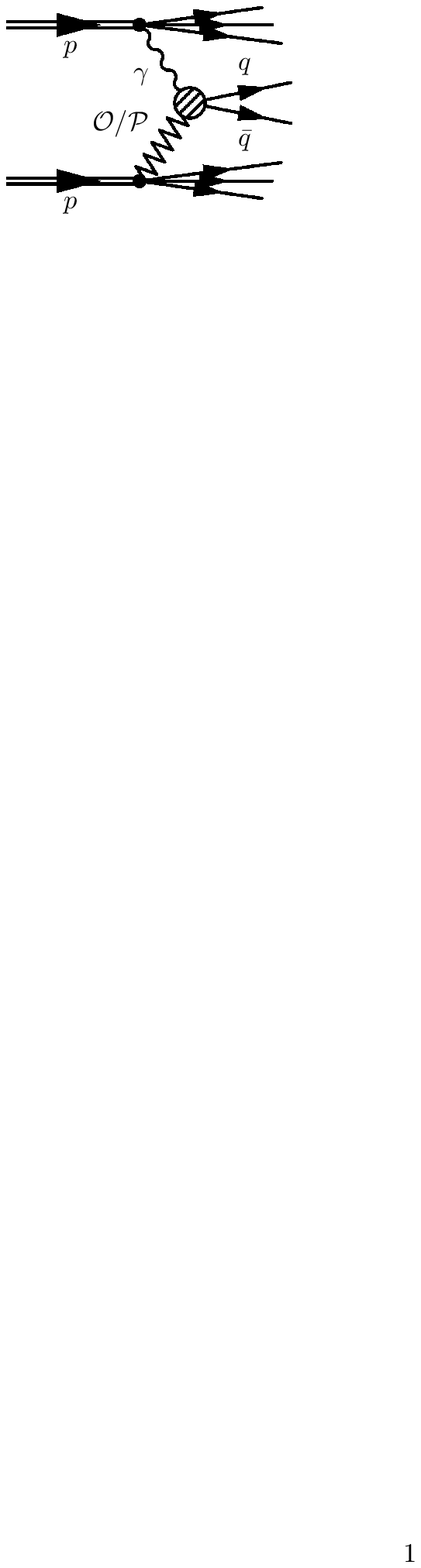}
\caption{photon-Pomeron and photon-Odderon amplitudes}
\label{fig:odderon_inter}
\end{center}
\end{figure}

Fig.\ref{fig:odderon_inter} shows the photon-Pomeron and the
photon-Odderon amplitudes for $q\bar{q}$ production.  
A study of open charm diffractive photoproduction estimates the asymmetry in 
fractional energy to be on the order of 15\%\cite{brodsky}. The 
forward-backward charge asymmetry in diffractive production of pion
pairs is calculated to be on the order of  10\% for pair masses in the 
range 
\mbox{$1\: GeV/c^{2} < m_{\pi+\pi-} < 1.3\: GeV/c^{2}$\cite{haegler,ginzburg}.} 

\section{Photoproduction of heavy quarks}
\label{sec:photo}

Diffractive reactions involve scattering on small-x gluons in the
proton. The number density  of gluons at given x increases with Q$^2$, 
as described by the DGLAP evolution. Here, Q$^2$ and x denote the
kinematical parameters used in deep inelastice ep scattering. The
transverse gluon density at a given Q$^2$ increases with decreasing x as
described by the BFKL evolution equation. At some density, gluons will
overlap and hence reinteract. In this regime, the gluon density
saturates and the linear DGLAP and BFKL equation reach their range of
applicability. A saturation scale Q$_s$(x) is defined which represents 
the breakdown of the linear regime. Nonlinear effects become visible
for Q $<$ Q$_s$(x). 

Diffractive heavy quark photoproduction represents an interesting probe to
look for gluon saturation effects at LHC. The inclusive cross section
for $Q\bar{Q}$ photoproduction can be calculated within the dipole
formalism. In this approach, the photon fluctuates into a $Q\bar{Q}$
excitation which interacts with the proton as a color dipole. The
dipole cross section $\sigma$(x,r) depends on x as well as on the
transverse distance r of the $Q\bar{Q}$ pair. A study of inclusive
heavy quark photoproduction in pp collisions at LHC energy has been 
carried out\cite{goncalves1}. 
These studies arrive at differential cross sections for open charm 
photoproduction of $\frac{d\sigma}{dy}\mid_{y=0} \; \sim$ 1.3 $\mu$b 
within the collinear pQCD approach as compared to 
$\frac{d\sigma}{dy}\mid_{y=0} \; \sim$ 0.4 $\mu$b within the color 
glass condensate (CGC). The cross sections are such that open charm  
photoproduction seems measurable with good statistical significance.
The corresponding numbers for the cross section for bottom 
photoproduction are $\frac{d\sigma}{dy}\mid_{y=0} \;
\sim$ 20 nb and 10 nb, respectively.

Diffractive photoproduction is characterized by two rapidity gaps in
the final state. In the dipole formalism described above, the two
gluons of the color dipole interaction are in color singlet state. 
Diffractive heavy quark photoproduction cross sections in 
pp, pPb and PbPb collisions at LHC have been studied\cite{goncalves2}.
The cross sections for diffractive charm photoproduction are 
$\frac{d\sigma}{dy}\mid_{y=0} \; \sim $ 6 nb in pp,
$\frac{d\sigma}{dy}\mid_{y=0} \; \sim $ 9 $\mu$b in pPb and
$\frac{d\sigma}{dy}\mid_{y=0} \; \sim $ 11 mb in PbPb collisions.
The corresponding numbers for diffractive bottom photoproduction are 
$\frac{d\sigma}{dy}\mid_{y=0} \; \sim $ 0.014 nb in pp,
$\frac{d\sigma}{dy}\mid_{y=0} \; \sim $ 0.016 $\mu$b in pPb and
$\frac{d\sigma}{dy}\mid_{y=0} \; \sim $ 0.02 mb in PbPb collisions.
 
Heavy quarks with two rapidity gaps in the final state can, however,
also be produced by central exclusive production, i.e. two Pomeron
fusion. The two production mechanisms have a different t-dependence. A
careful analysis of the transverse momentum $p_T$ of the $Q\bar{Q}$
pair might therefore allow to disentangle the two contributions.

\vspace{1cm}

{\bf Acknowledgments}

I  thank Otto Nachtmann and Carlo Ewerz for 
illuminating discussions and Leszek Motyka for preparing and
communicating Figure \ref{fig:odderon_pt}.




\begin{thebibliography}{29}

\bibitem{Alice1} F. Carminati et al, ALICE Collaboration, 2004, J.Phys. G:
  Nucl. Part. Phys. {\bf 30} 1517

\bibitem{Alice2} B. Alessandro et al, ALICE Collaboration, 2006, J.Phys. G:
  Nucl. Part. Phys. {\bf 32} 1295

\bibitem{ZDC}R. Arnaldi et al, Nucl. Instr. and Meth. A {\bf 564}
  (2006) 235

\bibitem{pixel} The ALICE collaboration, K. Aamodt et al., The ALICE
 experiment  at the CERN LHC, 2008\_JINST\_3\_S08002.

\bibitem{TOTEM} K. Eggert, TOTEM a different LHC experiment, CERN
  colloquium, feb 21,2008 

\bibitem{CMS} D. d'Enterria et al, Addendum CMS technical design report, 
J. Phys. G34:2307-2455, 2007

\bibitem{close}F. Close, A. Kirk, G. Schuler, Phys.Lett. B {\bf 477} (2000) 13

\bibitem{DL}A. Donnachie, P. Landshoff, Phys.Lett. B{\bf 595} (2004) 393 

\bibitem{akesson}T. \AA kesson et al, Nucl. Phys. B {\bf 264} (1986) 154 

\bibitem{nicolescu}L. Lukaszuk, B. Nicolescu, Lett. Nuovo Cim. {\bf 8}
  (1973) 406

\bibitem{ewerz}C. Ewerz, Proceedings XII Rencontres de Blois (2005) 377

\bibitem{pomeron_qcd} S. Donnachie, G. Dosch, P.V. Landshoff, O. Nachtmann,
Pomeron physics and QCD, Cambridge University Press (2002) 297

\bibitem{schaefer} A. Sch\"{a}fer, L. Mankiewicz, O. Nachtmann, 
Phys.Lett. B {\bf 272} (1991) 419

\bibitem{bzdak} A. Bzdak, L. Motyka, L. Szymanowski, J.R. Cudell, 
Phys.Rev. D {\bf 75} (2007) 094023

\bibitem{brodsky} S.J. Brodsky, J. Rathsman, C. Merino, 
Phys.Lett. B {\bf 461} (1999) 114

\bibitem{haegler} P. H\"{a}gler, B. Pire, L. Szymanowski, O.V. Teryaev,
Phys.Lett. B {\bf 535} (2002) 117 

\bibitem{ginzburg} I.F. Ginzburg, I.P. Ivanov, N.N. Nikolaev, 
Eur.Phys.J. C{\bf 5} (2003) 02

\bibitem{goncalves1}V.P. Goncalves, M.V. Machado,
Phys.Rev.D {\bf 71} (2005) 014025

\bibitem{goncalves2}V.P. Goncalves, M.V. Machado,
Phys.Rev.D {\bf 75} (2007) 031502


\end{thebibliography}
\end{document}